\documentclass[prl,10pt,twocolumn,superscriptaddress,showpacs]{revtex4}
\usepackage{amsmath}
\usepackage{amssymb}
\usepackage{bm}
\usepackage{graphicx}

\begin{document}

\title{New polarization effect and collective excitation in $S=1/2$
quasi 1D antiferromagnetic quantum spin chain}

\author{S.~V.~Demishev}
\author{A.~V.~Semeno}
\affiliation{General Physics Institute of Russian Academy of
Sciences, Vavilov str., 38, 119991 Moscow, Russia}
\author{H.~Ohta}
\author{S.~Okubo}
\affiliation{Molecular Photoscience Research Center, Kobe
University, 1-1 Rokkodai, Nada, Kobe 657-8501, Japan}
\author{I.~E.~Tarasenko}
\author{T.~V.~Ishchenko}
\author{N.~E.~Sluchanko}
\affiliation{General Physics Institute of Russian Academy of
Sciences, Vavilov str., 38, 119991 Moscow, Russia}

\begin{abstract}
Anomalous polarization characteristics of magnetic resonance in
CuGeO$_3$ doped with 2\% of Co impurity are reported. For the
Faraday geometry this mode is damped for the microwave field
$\mathbf{B}_\omega$ aligned along a certain crystallographic
direction showing that the character of magnetic oscillation differs
from the standard spin precession. The observed resonance coexists
with the EPR on Cu$^{2+}$ chains and argued not to be caused by
``impurity'' EPR, as previously claimed, but to correspond to an
unknown before, collective mode of magnetic oscillations in an
$S=1/2$ AF quantum spin chain.
\end{abstract}

\pacs{76.90.+d,75.30.Gw,75.45.+j}

\maketitle

From the modern theoretical point of view \cite{1,2} magnetic
resonances in $S=1/2$ quasi 1D antiferromagnetic (AF) quantum spin
chains should be treated as collective phenomena rather than a
diffusive spin dynamics suggested by exchange narrowing \cite{3}
approach. In the case of the electron paramagnetic resonance (EPR),
the collective nature develops in the specific temperature
dependences of the line width and $g$-factor caused by an interplay
of the staggered field and anisotropic exchange and in the damping
of EPR at low temperatures accompanied by rising of the breather
mode \cite{1,2}. In the latter case, the changes in the resonant
spectrum also affect the experiment geometry; namely, EPR can be
excited only in the Faraday geometry whereas the breather mode may
be observed for both the Faraday and Voigt geometry \cite{4}. These
predictions have been well proven experimentally \cite{4,5,6} for
the cases of Cu-benzoate and doped CuGeO$_3$.


Another possible field, where a collective motion of spins may be
found, is a polarization effect \cite{2}. However, although both the
field theory approach \cite{2} and direct numerical simulation
\cite{7} suggest that an EPR line depends on the orientation of the
microwave field $\mathbf{B}_\omega$, the expected influence on the
line width and $g$-factor is small. This result agrees well with the
previous calculations in the frame of exchange narrowing theory and
with known experimental data \cite{8,9,10}.

Here we report an experimental observation of a strong polarization
dependence for a magnetic resonance in CuGeO$_3$ doped with a Co
impurity, which have not been foreseen by existing theories for a
low dimensional magnets. We argue that the discovered effect
reflects the appearance of the unknown before collective mode in an
$S=1/2$ quasi 1D AF quantum spin chain.

A cobalt magnetic impurity in CuGeO$_3$ ($S=3/2$) substitutes cooper
in chains \cite{11,12,13} and in contrast to the other dopants
induces an onset of the specific resonant mode, which accompanies
EPR on Cu$^{2+}$ chains \cite{11,12}. Therefore, an experimental
spectrum of the resonant magnetoabsorption in CuGeO$_3$:Co is formed
by two broad lines, which can be completely resolved for frequencies
$\omega/2\pi\geq$100 GHz. It was found \cite{11,12} that frequencies
of both modes are proportional to the resonant magnetic field in a
wide range 60-360 GHz. The analysis of the $g$-factor values have
shown that the resonant mode corresponding to higher magnetic fields
represents a collective EPR on Cu$^{2+}$ chains, whereas the
resonant mode corresponding to lower magnetic fields may be
interpreted as an EPR on Co$^{2+}$ impurity clusters embedded into
CuGeO$_3$ matrix rather than as an antiferromagnetic resonance
(AFMR) mode \cite{12}.

In the present paper, we performed polarization measurements in the
Faraday geometry of the magnetic resonance spectrum of CuGeO$_3$
containing 2\% of Co at frequency 100 GHz in a temperature range
1.8-40 K. The details about samples preparation, characterization,
and quality control are given elsewhere \cite{12}. It was
established that for this concentration range Co impurity completely
damp the spin-Peierls transition for the vast majority of Cu$^{2+}$
chains and no Neel transition was found at least down to 1.8 K
\cite{11,12}. The quantitative analysis of the EPR on Cu$^{2+}$
chains parameters have shown that line width and $g$-factor reflect
properties of the chains with the damped spin-Peierls state and,
moreover, the Cu$^{2+}$ magnetic subsystem retains one dimensional
character in the aforementioned temperature interval \cite{5,12}.

In polarization experiments, a single crystal of \hbox{CuGeO$_3$:Co}
was located on one of the endplates of the cylindrical reflecting
cavity tuned to the TE$_{014}$ mode. A small DPPH reference sample
was simultaneously placed in the cavity. The external field
$\mathbf{B}$ up to \hbox{7 T} was generated by a superconducting
solenoid and was parallel to the cavity axis. Three cases, when
$\mathbf{B}$ was aligned along the $\mathbf{a}$, $\mathbf{b}$ and
$\mathbf{c}$ crystallographic directions, were studied. In each case
two orientations of the oscillating microwave field along remaining
axes were investigated; namely,  the polarizations
$\mathbf{B}_\omega\Vert\mathbf{a}$ and
$\mathbf{B}_\omega\Vert\mathbf{c}$ for $\mathbf{B}\Vert\mathbf{a}$
and so on. (Hereafter we denote the Cu$^{2+}$ chain direction as
$\mathbf{c}$; $\mathbf{b}$ axis is perpendicular to $\mathbf{c}$
marking a direction of the second strongest exchange in the chains
plane and $\mathbf{a}$ axis is orthogonal to the chains plane).
Therefore below we report the results for six experiment geometries.
Measurements were repeated for ten single crystals and provided
identical results.

It is worth noting that $\mathbf{B}_\omega\perp\mathbf{B}$ in all
cases studied and no effect in a convenient EPR on a single spin is
expected. For the isotropic Heisenberg spin system, the spin
Hamiltonian consisting of the exchange and Zeeman terms commutes
with the magnitude of the total spin and its $z$-component. Thus in
the absence of anisotropic terms in the Hamiltonian, which give rise
to a finite line width, the EPR occurs at the same frequency
$\omega=\gamma B$ as in single-spin problem \cite{1,2}. As a result,
even in a strongly interacting system like quantum spin chain, it is
possible to use a semiclassical language and describe a magnetic
resonance in terms of magnetization rotation along the field
direction as one does for the single spin. Therefore for the quantum
spin chain in ``zero order'', the excitation of EPR does not depend
on the $\mathbf{B}_\omega$ direction in plane perpendicular to
$\mathbf{B}$. This speculation qualitatively explains why
polarization effects may be only expected in the line width and
polarization corrections to the $g$-factor that is in agreement with
the exact results \cite{2,7,8,9}.

The obtained experimental data, however, contradict to this picture.
As can be seen from Fig. 1 the low field mode A, which is suspected
to be an ``impurity'' resonance in the previous studies
\cite{11,12}, can be excited for one polarization only. At the same
time, the EPR on Cu$^{2+}$ chains (resonance B) does not show any
strong polarization dependence. For the mode A and
$\mathbf{B}\Vert\mathbf{a}$, an ``active'' polarization is
$\mathbf{B}_\omega\Vert\mathbf{c}$ and ``non-active'' polarization
corresponds to the case $\mathbf{B}_\omega\Vert\mathbf{b}$ (Fig. 1,
a). It is worth to note that in the case of ``non-active''
polarization the magnetic resonance A is almost completely damped,
and weak traces of this mode for $\mathbf{B}_\omega\Vert\mathbf{b}$,
which are visible at low temperatures, are due to the finite sample
size and related weak misalignment of $\mathbf{B}_\omega$ from
$\mathbf{b}$ axis in cavity measurements. Another characteristic
feature of the observed phenomenon is the peculiar temperature
dependence. For the ``active'' case, mode A appears below 40 K and
at $T\sim$12 K becomes as strong as the resonance on Cu$^{2+}$
chains. Further lowering of temperature makes the A resonance a
dominating feature in the magnetoabsorption spectrum with amplitude
considerably exceeding that of mode B (Fig. 1, b).

\begin{figure}

\includegraphics[width=0.8\linewidth]{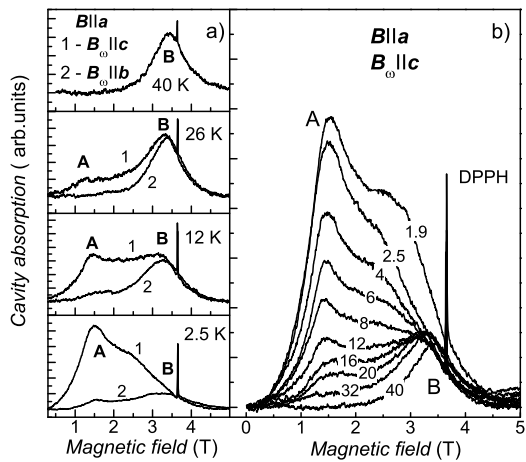}

\caption{Comparison of ``active'' and ``non-active'' polarizations
(panel a) and temperature evolution of the magnetoabsorption
spectrum for ``active'' polarization (panel b) in geometry
$\mathbf{B}\Vert\mathbf{a}$. Narrow line in panel a represents DPPH
signal. Figures near curves in panel b correspond to temperatures in
K.} \label{fig:1}

\end{figure}

Similar behavior is observed for $\mathbf{B}\Vert\mathbf{b}$
geometry (Fig. 2). In this case for mode A an ``active''
polarization is $\mathbf{B}_\omega\Vert\mathbf{a}$, and
``non-active'' polarization is $\mathbf{B}_\omega\Vert\mathbf{c}$,
whereas the resonance B is not much affected by an orientation of
the microwave field. In agreement with the case
$\mathbf{B}\Vert\mathbf{a}$, the mode A is the strongest in the
spectrum, however for $\mathbf{B}\Vert\mathbf{b}$ the main resonance
A is accompanied by its second harmonic (Fig. 2). Interesting that
although the mode A is completely damped in
$\mathbf{B}_\omega\Vert\mathbf{c}$ case, the second harmonic of this
resonance retains the same amplitude for both ``active'' and
``non-active'' polarizations (Fig. 2).

\begin{figure}

\includegraphics[width=0.8\linewidth]{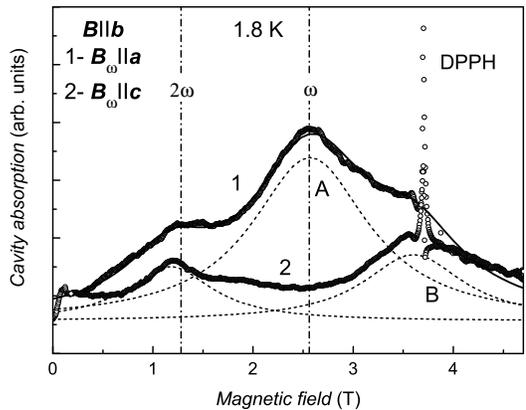}

\caption{``Active'' and ``non-active'' polarizations for
$\mathbf{B}\Vert\mathbf{b}$. Points correspond to experiment, solid
line- fitting of the experimental spectrum assuming Lorentzian
shapes of the resonances A, B and second harmonic of resonance A.
Partial contributions of these resonances to the spectrum are shown
by dashed lines.} \label{fig:2}

\end{figure}

A dominating character of the resonance A at low temperatures is
conserved in $\mathbf{B}\Vert\mathbf{c}$ case (Fig. 3). The effect
of polarization appears to be weaker, and  the amplitude of the
resonance A for $\mathbf{B}_\omega\Vert\mathbf{b}$ is only two times
less than for $\mathbf{B}_\omega\Vert\mathbf{a}$. Nevertheless, the
polarization dependence of this mode remains anomalously strong,
especially as compared with the resonance on Cu$^{2+}$ chains.

\begin{figure}

\includegraphics[width=0.8\linewidth]{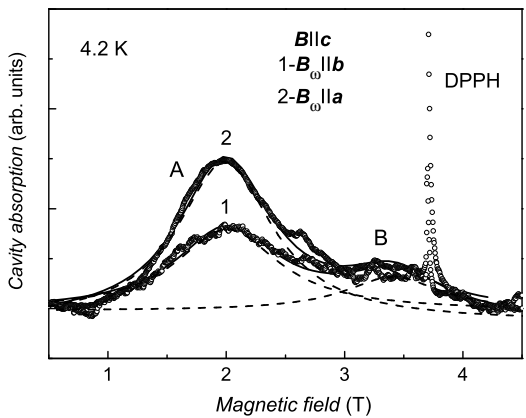}

\caption{``Active'' and ``non-active'' polarizations for
$\mathbf{B}\Vert\mathbf{c}$. Points correspond to the experiment,
solid line- to fitting of the experimental spectrum assuming
Lorentzian shapes of the resonances A and B. Partial contributions
of these resonances to the spectrum are shown by dashed lines.}
\label{fig:3}

\end{figure}

Data in Figs. 1-3 show that the resonance field for the mode A
varies substantially, when a direction of the external magnetic
field $\mathbf{B}$ is changed. The corresponding $g$-factor values
are $g\approx$4.9 ($\mathbf{B}\Vert\mathbf{a}$), $g\approx$2.9
($\mathbf{B}\Vert\mathbf{b}$), and $g\approx$3.7
($\mathbf{B}\Vert\mathbf{c}$). Thus, the $g$-factor for this mode
may differ 1.7 times, while for the EPR on Cu$^{2+}$ chains
(resonance B) the $g$-factors for various crystallographic
directions lie in the range 2.06-2.26 \cite{6,12} and, hence, are
changed only by 10\%.

The experimental data obtained in the present work demonstrate that
the resonant mode A in CuGeO$_3$:Co is an anomalous one. First of
all, this mode shows extremely strong dependence on orientation of
the oscillating microwave field $\mathbf{B}_\omega$ in the Faraday
geometry. At the same time, no comparable effect for EPR on
Cu$^{2+}$ chains is observed in good agreement with theoretical
expectations \cite{7,8,9,10}.

Secondly, the vanishing of the resonance A for certain polarizations
means that the character of magnetic oscillations in this mode is
completely different from the magnetization vector precession around
a magnetic field direction described by Landau-Lifshits equation.
Indeed, in case of precession, the magnetization vector end moves
around a circle, whose plane is perpendicular to the magnetic field,
and hence any linear polarization in Faraday geometry excites an
EPR-like mode or a mode based on correlated precession of various
magnetization components \cite{14}.

Thirdly, modes A and B coexist in a wide temperature range $1.8<T<
40$ K and therefore the scenario giving a vanishing collective EPR
(mode B) after an onset of the breather excitation like in
Cu-benzoate \cite{1,2,4} does not hold. The appearance of the mode A
at relatively high temperatures $T\sim$40 K (Fig. 1) simultaneously
eliminates applicability of the standard scenario of doping
\cite{15}, where the coexistence of EPR and AFMR in doped CuGeO$_3$
may be expected only at temperatures below $0.3 T_{SP}\sim$4 K (AFMR
coexisting with EPR in CuGeO$_3$ have been observed in experiments
at $T<$2 K \cite{16}).

The above consideration does not allow explaining of mode A in terms
of a single spin EPR problem. At the same time, the properties of
this magnetic resonance is not possible to describe assuming either
collective EPR or AFMR in a quantum spin chain system, as well by
the other known to date collective modes like breather excitations.
Thus, the doping with Co of Cu$^{2+}$ quantum spin chains in
CuGeO$_3$ leads to formation of a novel unidentified magnetic
resonance. Nevertheless it is possible to deduce that the observed
excitation of a magnetic subsystem of CuGeO$_3$:Co has a collective
nature. The first argument favoring this supposition is the
magnitude of this magnetic resonance. Taking into account that in
the samples studied only 2\% of cooper ions are substituted by the
cobalt impurity and no spin-Peierls transition affecting mode B
happens, it is difficult to expect that any individual impurity mode
considerably exceeds the magnitude of the magnetic resonance on
Cu$^{2+}$ chains (Figs. 1-3). Therefore, in our opinion, mode A is
likely a specific collective excitation of a quasi 1D Cu$^{2+}$
chain, which properties are modified by doping.

The unusual polarization dependence of the mode A may be considered
as another argument. Apparently, the observed behavior is forbidden
for a single spin or $S=1/2$ AF spin chain with an isotropic
Hamiltonian \cite{1,2}. However, in presence of anisotropic terms,
the spin chain Hamiltonian does not commute with the total spin and
its $z$-component, and hence, in principle, the magnetic oscillation
modes different from the standard spin precession may become
possible. It is worth to note that experimental data in Figs. 1-3
suggest a selected character of $\mathbf{b}$ axis. Indeed, for
$\mathbf{B}_\omega\Vert\mathbf{b}$ the resonance A is completely
damped ($\mathbf{B}\Vert\mathbf{a}$) or its magnitude is reduced
($\mathbf{B}\Vert\mathbf{c}$) and in case
$\mathbf{B}\Vert\mathbf{b}$ the second harmonic of the anomalous
mode, which is missing in other geometries, develops. At the same
time, the previous studies \cite{5,6} have shown that doping with
magnetic impurities of CuGeO$_3$ leads to appearance of the
staggered magnetization aligned predominantly along $\mathbf{b}$
axis \cite{6}. Therefore, the observed mode A is likely related with
the staggered field, which may be responsible for anomalous
polarization characteristics. Apparently no such effects caused by a
staggered field may be expected for a single spin resonance and thus
the idea of a staggered magnetization controlled mode A agrees with
its collective nature. From to date theoretical point of view, a
staggered field is known to be anisotropic term in Hamiltonian,
which is crucial for the EPR problem in the studied case \cite{1,2}.
However the question whether this type of anisotropy is sufficient
for the explanation of the observed phenomena remains open and a
required extension of the theory \cite{1,2} is missing.

From the data presented in Figs. 1-3 it is possible to deduce the
character of magnetic oscillations for resonances A and B in
CuGeO$_3$:Co. In a semiclassical approximation the magnetization in
a given magnetic field $\mathbf{B}$ has the form
$\mathbf{M}=\mathbf{M}_0+\mathbf{m}$, where $\mathbf{M}_0$ denotes
an equilibrium value and $\mathbf{m}$ stands for the oscillating
part \cite{14}. As long as magnetic resonances probe normal modes of
magnetization oscillations described by vector $\mathbf{m}$, for an
excitation of some mode, vector $\mathbf{B}_\omega$ should have non
zero projection on any $\mathbf{m}$ component \cite{14}, i.e., a
condition $(\mathbf{B}_\omega,\mathbf{m})\neq0$ for the scalar
product should be fulfilled. For geometry
$\mathbf{B}\Vert\mathbf{a}$ and a normal mode, when precession of
magnetization around the field direction takes place, $\mathbf{m}
=(0,{m}_b,{m}_c)$ and both projections of $\mathbf{m}$ on
$\mathbf{b}$ and $\mathbf{c}$ axes are nonzero. Therefore any
alignment of vector $\mathbf{B}_\omega$ in $\mathbf{b}\mathbf{c}$
plane excites precession. The weak dependence of the resonance
amplitude on $\mathbf{B}_\omega$ alignment corresponds to condition
${m}_b\approx{m}_c$. Thus, for the mode B and
$\mathbf{B}\Vert\mathbf{a}$,  the trajectory of the vector
$\mathbf{M}$ end is a circle lying in $\mathbf{b}\mathbf{c}$ plane
(a similar consideration is apparently applicable to mode B in
geometries $\mathbf{B}\Vert\mathbf{b}$ and
$\mathbf{B}\Vert\mathbf{c}$).

The same analysis can be applied for the resonance A. Data in Fig. 1
suggest that in geometry $\mathbf{B}\Vert\mathbf{a}$ the oscillating
contribution to magnetization should acquire the form $\mathbf{m}
=(0,0,{m}_c)$ leading to ``active'' polarization
$\mathbf{B}_\omega\Vert\mathbf{c}$ and ``non-active'' polarization
$\mathbf{B}_\omega\Vert\mathbf{b}$ (Fig. 1). Therefore in this case,
the end of vector $\mathbf{M}$ should move along the line parallel
to $\mathbf{c}$ axis. Analogously, $\mathbf{m} =({m}_a,0,0)$ for
$\mathbf{B}\Vert\mathbf{b}$ and linear oscillation happens along
$\mathbf{a}$ axis (Fig. 2). For $\mathbf{B}\Vert\mathbf{c}$, mode A
can be excited in both polarizations and hence $\mathbf{m}
=({m}_a,{m}_b,0)$. However, the decrease of the resonance magnitude
for $\mathbf{B}_\omega\Vert\mathbf{b}$  suggests condition
${m}_a\approx2{m}_b$ (Fig. 3). As a result, the trajectory of the
vector $\mathbf{M}$ end will be an ellipse in $\mathbf{a}\mathbf{b}$
plane elongated in $\mathbf{a}$ direction. The summary of the above
consideration for mode A is given in Fig. 4.

\begin{figure}

\includegraphics[width=0.6\linewidth]{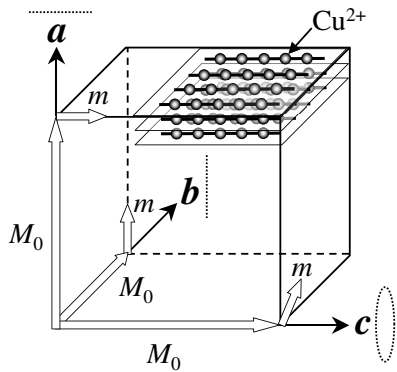}

\caption{Schema of possible magnetic oscillations for normal modes
responsible for anomalous polarization dependences of the mode A in
three cases ($\mathbf{B}\Vert\mathbf{M}_0\Vert\mathbf{a}$,
$\mathbf{B}\Vert\mathbf{M}_0\Vert\mathbf{b}$ and
$\mathbf{B}\Vert\mathbf{M}_0\Vert\mathbf{c}$). Oscillating
contribution $\mathbf{m}$ is assumed to vary harmonically with time
in cases $\mathbf{M}_0\Vert\mathbf{a}$ and
$\mathbf{M}_0\Vert\mathbf{b}$. For $\mathbf{M}_0\Vert\mathbf{c}$
vector $\mathbf{m}$ rotates around the $\mathbf{c}$ axis. Dotted
lines mark trajectories of the vector
$\mathbf{M}=\mathbf{M}_0+\mathbf{m}$ end.} \label{fig:4}

\end{figure}

To our best knowledge, the modes with linear or elliptic oscillation
trajectories have been neither reported for any magnetic resonance,
nor foreseen by theoretical studies. Moreover, the current
understanding of the whole field of magnetic resonance (including
EPR, AFMR and ferromagnetic resonance) essentially exploits
semiclassical magnetization precession in an external field, and
hence leaves no space to the observed new polarization effect.
Therefore, an adequate theory relevant to the studied case,
including different polarization characteristics of magnetic
resonance harmonics appears on the agenda.

In conclusion, we have shown that doping of CuGeO$_3$ with 2\% of Co
impurity induces an anomalous magnetic resonance mode possessing
unique polarization characteristics. This resonance coexists with
the EPR on Cu$^{2+}$ chains and is likely caused by a new, unknown
before, collective mode of magnetic oscillations in $S=1/2$ AF
quantum spin chain.

\begin{acknowledgments}
Authors are grateful to M.Oshikawa for the stimulating discussion.
This work was supported by the RFBR grant 04-02-16574 and by the
Programme "Strongly correlated electrons" of Russian Academy of
Sciences.
\end{acknowledgments}

\end{document}